\documentclass[]{mn2e}
\usepackage[dvips]{graphicx} 
\title[Observing Baryonic Dark Matter]
{Observing Baryonic Dark Matter with ALMA}
\author[H.Kamaya and J.Silk]
{Hideyuki Kamaya$^{1,2}$\thanks{Email:kamaya@kusastro.kyoto-u.ac.jp}
and Joseph Silk$^{1}$ \\
$^{1}$ Astrophysics, Department of Physics, Denys Wilkinson Building, 
University of Oxford, Oxford OX1 3RH \\
$^{2}$ Department of Astronomy, School of Science,  
Kyoto University, Kyoto, 606-8502, Japan
}

\begin{document}
\maketitle 
\begin{abstract}

It has recently been argued that the unidentified SCUBA objects (USOs)
are a thick disk population of free-floating dense, compact galactic
gas clumps at a temperature of about 7 K. The characteristic mass
scale is constrained to be on the order of a Jupiter mass, and the
size is about 10 AU. A typical galactic USO is located at a distance
from the sun of about 300 pc.  We have calculated the molecular
emission lines from these low temperature clouds. We consider three
molecules: HD, LiH, and CO. HD is optically thin in the cloud, LiH is
a molecule with a large electric dipole moment, and CO is an
abundant molecule that is observed in dusty clouds.  Our estimate for
the typical object shows that LiH may be  detectable by the future sub-mm
array project, ALMA; its expected flux is at the mJy level and the
line width is about $10^5$ Hz. Although typical galactic USOs are
chemically and dynamically transient, the younger USOs will be
recognisable via LiH emission if about a hundred USOs are observed.
If USOs are confirmed to be of galactic origin, the total baryonic
budget will need to  be reevaluated.

\end{abstract}


\section{INTRODUCTION}

Standard cosmic nucleosynthesis calculations together with the current
observational measurements of the abundance of $^2$D, in addition to
limits on $^3$He, $^4$He, and $^7$Li, constrain the cosmic baryon
density to be (O'Meara et al. 2001)
$$
\Omega_{\rm b} h^2 = 0.02\pm 0.002. 
\eqno(1) $$
The Hubble constant, $H_0 = 100 h$ km s$^{-1}$ Mpc$^{-1}$, is
estimated to be $h=0.72 \pm 0.08$ (Freedman et al. 2001).  Hence the
cosmic baryon density is between 0.03 and 0.05.  There is a similar
constraint on $\Omega_{\rm b} h^2$ from the CMB (e.g. Ferreira 2001;
Bartlett 2001 for reviews), the most recent result being $\Omega_{\rm
b} h^2 = 0.033 \pm 0.013$ (Stompor et al. 2001).  Stringent
constraints from the CMB anisotropies limit
 the extreme possibility that
baryons account for all of the matter in the universe 
(Griffiths, Melchiorri, \& Silk (2001).

In fact, the baryon budget is poorly known (cf.  Persic \& Salucci
1992; Gnedin \& Ostriker 1992; Bristow \& Phillipps 1994; Fukugita,
Hogan \& Peebles 1998). Table 3 of Fukugita et al. summarises the
results.  While the stellar contributions are minor, hot plasma in
clusters and groups of galaxies is the dominant contributor to the
observed baryon budget.  However, for example, the contribution from
the intracluster medium is considered to be only in the form of hot
plasma. Any warm gas or small, cold clouds in the clusters would
contribute to the baryon budget.  While the conclusion of Fukugita et
al. is consistent with $\Omega_{\rm b}$ as constrained above, the
possibility remains that the contribution of small cold clouds is
significant.  This has indeed been argued to be the case for the
intracluster medium, but a stronger case can perhaps be made for the
presence of such clouds in less hot environments, such as the
intra-group medium and galaxy halos, where observations are less
constraining.  For example, if there is undetected galactic HI
and CO, the galactic baryon budget in the form of tiny  very cold
clouds may be underestimated.

Possible dynamical structures of such  tiny very cold clouds have been
 examined by Gerhard \& Silk (1996). According to these authors, in order to
stabilize these  low temperature clouds against collapse, two mechanisms are
of interest. One  is gas-dust collisional heating, and the
other is the effect of the gravitational potential wells of
collisionless dark matter. Other aspects are discussed in many papers 
(e.g. de Paolis et al. 1995a, 1995b; Henriksen \& Widrow 1995, Draine
1998; Kalberla et al. 1999; Wardle \& Walker 1999).  In this $Letter$,
we also comment on the dynamical properties very briefly in the next
section, focussing on the  observational claim that some very 
low temperature clouds may have retained or acquired dust (e.g. Lawrence 2001).

Indeed, it has been suggested observationally that halo dark matter
could reside in the form of cold dark clouds and so be
undetected. Pfenniger \& Combes (1994) and Pfenniger et al. (1994)
argued that such clouds would be at or near the traditional
hierarchical fragmentation limit and further argued that flat rotation
curves could be explained by baryonic dark matter in this form. Walker
\& Wardle (1998) examined observational constraints in the radio bands for
the clouds. The observational possiblity of an emission line forest (i.e.
quasi-continuum) from the clouds was first examined by Sciama (2000).
An extensive  review of the small--scale interstellar medium, which may be 
related to these tiny cold clouds, is presented in Heiles (1997). In
the same context, Lawrence (2001) has proposed a  possible galactic
origin for  the unidentified SCUBA objects (USOs).  In order for the
USOs to be galactic, the far IR/mm background and SCUBA source counts
at 400 and 800 $\mu$m, together with dynamical limits on galactic dark
matter, constrain the cloud parameters, if distributed throughout the
galaxy, to have low temperatures $\sim 7$K, Jupiter-like masses $M_{\rm
J}$, and to be very small, about 10 AU in size. As long as these
characteristics are imposeed because of the dynamical limits, it may
be  that USOs contribute to the baryon budget in the
intracluster and intra-group medium although they  have a small
covering factor over the sky. In this $Letter$, we propose a simple
observational strategy for the direct detection of galactic USOs.

\section{Physical Condition of Unidentified SCUBA Objects}

The physical conditions in USOs may be summarised as follows.
Lawrence (2001) posed a number of constraints on the sub-mm
sources. According to his discussion: (i) the objects must have an SED
consistent with what is known from the brighter sub-mm sources; (ii)
their distribution on the sky must be at least roughly isotropic,
although current surveys cannot exclude variations of the order of
tens of percent; (iii) the deduced space density must not exceed the
limits imposed by local dynamics; (iv) their integrated surface
brightness must not exceed the FIR-mm background discovered by COBE;
(v) the population covering factor must be small, or the objects would
already have been discovered through extinction effects.  From these
constraints, Lawrence developed a possible model for galactic SCUBA
point-sources, summarised in table 1 of his paper. We have extracted
three representative models in our table 1.  Jupiter--mass objects
satisfy all of the above criteria. Radio and sub-mm fluxes and upper
limits constrain the galactic USOs to have a temperature of about 7 K. 
We note that the SCUBA sources show continuous emission in the sub-mm
band, and hence they must retain some dust.  In a related paper, it is
shown that the USO covering factor is quantitatively constrained by
occultation limits, using data from  microlensing experiments
(Kerins, Binney and Silk 2002).

Possible dynamical properties of USOs are also of interest. Here, we
review very briefly the suggestions of Gerhard \& Silk (1996). The
expected low temperature clouds in the galactic halo, if primordial,
could also be maintained in the gravitational potential wells of
mini-clusters of collisionless dark matter.  Following Umemura \&
Ikeuchi (1986), Gerhard \& Silk found that the tendency to
self-gravitational fragmentation of the very low temperature clouds is
possibly stabilised by the gravity of the minicluster of collisionless
dark matter. The maximum mass fraction of the cold gas relative to
collisionless dark matter is expected to be about 0.01 -- 0.1. Thus, the
total mass of a typical USO as discussed by Lawrence could be as large
as 0.01 -- 1.0 solar mass without breaking the universal collisionless
cold dark matter assumption. Unfortunately, however, the current model
requires USOs to have dust, so that USOs are unlikely to be primordial.
Of course they may have accreted substantial amounts of dust in
traversing the galactic disk (Kerins et al. 2002).  Another means of
avoiding gravitational contraction and star or planet formation is
possible if the heating of the gas is dominated by gas-dust collisions
and the dust temperature is much higher than that of the gas (Gerhard \&
Silk 1996). This effect can stabilize the clouds for the age of the
Galaxy, since the effective specific heat ratio becomes larger than 4/3. 
To check this possibility, the dynamical coupling between dust and cold
gas is examined in a later section.

\section{Molecule Selection}

We examine the possibility of observing line emission from galactic
USOs.  The USOs are very low temperature clouds, and so  some molecular
emission is expected in the radio and sub-mm bands. The USOs consist
primarily of hydrogen molecules. Unfortunately, since the first level
for rotational excitations of H$_2$ is at 515 K, the expected line
flux of H$_2$ is not large. Hence, Sciama (2000) argues that there is
a quasi-continuum of many lines of hydrogen and heavy molecules (H$_2$,
CO, H$_2$O, HCl, O$_2$) to obtain the cooling rate of gas in the clouds
proposed by Walker and Wardle (1998). In this $Letter$, we further
consider three typical emission lines for  the following reasons.

The first molecule we examine is HD. This molecule is selected since it
should exist when the expected small cold cloud is composed mainly of
H$_2$. The HD molecule is generally optically thin in the clouds since
HD has a small electric dipole moment.  This moment has been measured in
the ground vibrational state from the intensity of the pure rotational
spectrum by Trefler \& Gush (1968). The measured value is about
5.85$\times 10^{-4}$ Debye (1 Debye $=$ $10^{-18}$ in cgs units).  Since
the first rotational level is about 128 K, the corresponding wavelength
is 112 $\mu $m.  In this paper, we set the D abundance to be $4\times
10^{-5}$ in number and assume that all the D is in the form of HD.  The
fraction of rotationally excited HD is determined for a thermal
distribution.

The second molecule we consider is LiH, which has a large electric
dipole moment of 5.9 Debyes. Hence the detection of LiH is generally of
interest for interstellar astrophysics. Indeed, there is a real
possibility that LiH is an important emission source (Lepp \& Shull
1984) because of its large electric dipole moment. Furthermore, the
detection of LiH itself is interesting because the surface conditions of
the (pre-Galactic) USOs must be linked with the pre-Galactic cosmic-ray
particle flux. This is because Li has the unique property of being
produced not only by big bang nucleosynthesis, but also by spallation of
Galactic cosmic-ray particles on interstellar matter nuclei (of course,
stellar nucleosynthesis is also important). Its abundance is expected to
be very small (e.g. $\sim 10^{-10}$ in number; we assume this value for
a typical case), although the variance of the measured abundance of Li
is known to be large (e.g. Hill, Barbuy, \& Spite 1997, Vangioni-Flam,
Coc, \& Cass\'e 2000, Travaglio et al. 2001).  The first rotational
level of LiH is only at about 21 K and its rest wavelength is about 0.67
mm.  For the case of LiH, we also adopt a thermal distribution.

The third molecule we consider is CO. This is the most common heavy
molecule in mm astronomy of the interstellar medium, which is why we
select it.  Comparison of the abundance of CO to that of LiH may provide
useful information on the evolution and formation of USOs. The USOs have
acquired dust, and so would also presumably have the C and O to form CO
in situ.  The CO dipole moment is 0.112 Debye, and the first rotational
level is about 5.5 K, corresponding to 2.6 mm in the rest frame. We
assume the CO/H ratio to be 1.0 $\times 10^{-4}$ which is about that of
the solar neighbourhood and consistent with the model proposed in
Lawrence (2001).  To determine the abundance in the $J=1$ level of CO,
we adopt a thermal distribution.  We note that the level population of
CO up to $J=3$ can be thermally excited even at a temperature of 7 K.

\section{Results}

To obtain the luminosity in the emission lines, we need to specify the
cloud volume, optical depth, and critical density for collisional
de-excitations to be dominant.  Fortunately, a  galactic USO is very
dense by definition (see table 1).  For all of the ground rotational
emission lines of the three molecules, the gas density of 0.1$M_{\rm
J}$ and 1.0$M_{\rm J}$ clouds is much larger than the critical density. 
The density of the 10 $M_{\rm J}$ example  is comparable to the critical
density of the ground rotational emission of LiH, while it is much
larger than the critical density for  HD and CO.  In these 
situations, to estimate the luminosity of the emission lines, we need
the de-excitation probability, $A_{J'J}$, and the optical depth,
$\tau_{\nu_{J'J}}$. We adopt a useful analytical formula for these
quantities:
$$ 
A_{J'J} = \frac{64\pi^4\nu_{J'J}^3}{3hc^3}D^2\frac{J}{2J+1}
\eqno(2)$$
and
$$
\tau_{\nu_{J'J}} = \frac{A_{J'J} c^3}{8 \pi \nu^3_{J'J}} 
\left(\frac{g_{J'}}{g_J} - \frac{n_{J'}}{n_{J}}\right) n_J 
\frac{R_{\rm J}}{\delta v}  , 
\eqno(3)$$ 
where $\nu_{J'J}$ is the frequency of the $J'\to J $ transition, $h$
the Planck constant, $c$ the light speed, $D$ the electric dipole
moment, $J$ the rotational quantum number, $R_{\rm J}$ the cloud
radius, $\delta v$ the velocity dispersion, and $g_J$ the statistical
weight of $2J+1$.  The velocity dispersion corresponds to Doppler
broadening, and is taken to be equivalent to the sound speed at 7K.
The effect of the turbulence and rotation of an USO can contribute to
$\delta v$. USOs are assumed to be virial equilibrium, and then the effect
of the turbulence and the rotation should be the same order as that of 
the thermal motions. Hence, the uncertainty of $\delta v$ owing to
cloud  turbulence and  rotation is a factor of a few.

The clouds are very cold, and so we consider mainly the first
rotational level of each molecule as a typical case. With (3), the
escape probability, $\epsilon_{\nu_{J'J}}$, of the photons becomes
$$
\epsilon_{\nu_{J'J}} 
= \frac{1 - {\rm exp}(\tau_{\nu_{J'J}}) }{\tau_{\nu_{J'J}}}. 
\eqno(4)$$ 
Finally, we can estimate the emission line as $n_{i,J'}E_{J',J}A_{J'J}
\epsilon_{\nu_{J'J}}$ erg cm$^{-3}$ sec$^{-1}$, where $n_{i,J'}$ is
the number density of the $i$-molecule with rotational level $J'$ and
$E_{J',J}$ is the transition energy between $J'$ and $J$. This formula
is reasonable as long as the net gas density is above any relevant
critical density and the temperature is very low.  For simplicity, we
assume that the cloud is uniform.

The expected fluxes at the Earth are summarised in table 1 for each 
model of the very low temperature cloud of Lawrence (i.e. for a 
typical galactic USO). Our results are the following.  

(1) HD: the three flux levels are similar to each other.  This is
because HD is optically thin and the cloud model of Lawrence is well
constrained by the other observational flux limits.  That is, the size
of the cloud and the typical distance from us is specified.  These
results for HD confirm that our estimates are reasonable.  Since the
flux level of HD is so low, it will be very difficult to detect HD
from the galactic USOs directly.

(2) LiH: we find this molecule to be very important.  The flux level
is at the mJy level, which should be observable by
ALMA\footnote{http:// www.alma.nrao.edu}.  ALMA is a ground-based
radio interferometric facility, and will consist of 96 12-m antennas.
A detailed recent review is found in Takeuchi et al. (2001), from
which we infer that the 5$\sigma $ sensitivities at 350 $\mu$m, 450,
650, 850, 1.3 mm, 3.0 mm are expected to be 390, 220, 120, 16, 7.5,
4.6 $\mu$Jy, respectively (8-GHz bandwidth).  Obviously, the predicted
frequency of LiH is located in the range of ALMA.  The typical line
width is about $10^5$ Hz.  The frequency resolution limit of ALMA is
about $4 \times 10^4$ Hz.  {\it Direct detection of LiH from galactic
USOs should be possible by ALMA.}

(3) CO : the expected flux level seems to be marginal for current
observational facilities, while the width of the line is very narrow
because of the low temperature.  Then, it is very reasonable to infer
that these very cold clouds would not yet been observed. However
follow-up CO detection is important if LiH is detected. This is
because we can confirm temperature and density of the USOs.

To predict the expected signal-to-noise ratio, it is necessary to know
the continuum level of the flux. We estimate this under the assumption
that the dust emits optically thick blackbody radiation. For the typical
case, the dust number density, $n_{\rm dust}$, is 18.5 cm$^{-3}$ and the
mean surface area is about $1.5 \times 10^{-11}$ cm$^{2}$, both of which
are determined under the assumption of a standard MRN number
distribution function (Mathis, Rumpl, \& Nordsieck 1977) with the
constraint that the total dust mass is $0.01 \times $ total mass of the
cold cloud, taken to be $M_J$ for a standard galactic USO. The specific
density of dust is assumed to be 2.0 g$\,$cm$^{-3}$, and the range of
dust radii, $a$, is adopted to be $0.0001$ -- 3.0$\times 10^{-7}$ cm.
Since the size of the cloud is 1 -- 100 AU, then we estimate the dust
continuum to be from an optically thick dust sphere with a typical
radius corresponding to each dust model.  Each estimated flux is
presented in table 1 for each of the wavelengths of HD, LiH, and CO. A
schematic view is also presented in figure 1 with expected line emission
for the typical case of $M_J$.

\section{Discussion} 

In the previous sections, USOs are assumed to have a constant abundance 
of molecules. However, USOs are cold and dense. In such objects, molecules 
stick to dust and are depleted from the gas phase. Then, we must examine
the depletion of LiH and CO. This is necessary if we try to find observational
feasibility. Unfortunately, the depletion of LiH is not well understood, 
and so we start by considering the case of CO. 

According to Duley \& Smith (1995), the sticking probability, $S_{\rm
co}$, of CO on dust is about 0.037 -- 0.0037. They derived this from
observations of heavy reddened regions in the Taurus molecular
clouds. For order of magnitude estimates, we shall adopt 0.01 as
$S_{\rm co}$. The exponential depletion rate is defined by $S_{\rm co}
\pi a^2 n_{\rm dust} v_{\rm s}$, where $ v_{\rm s}$ is the sound speed
of the gas around the dust. For $a$ and $n_{\rm dust}$, the
characteristic values for the MRN distribution are assumed.  Then, the
depletion rate is $0.3 \times 10^{-7}$ s$^{-1}$ with $v_{\rm s} = 1.13
\times 10^4$ cm s$^{-1}$ for the case of $M_{\rm J}$.

This is rather large, but of course we also need to specify the rate
at which cosmic rays and ambient photons may eject molecules from the
grain surfaces.  This is poorly known, but unlikely to be dominant in
these very dense clouds. If CO were to be observed from a typical
galactic USO, it would most likely be transient, and it would be
difficult to confirm the predicted universality of USOs.

The order of magnitude of the sticking probability of LiH, $S_{\rm
lih}$, may not be much smaller than $S_{\rm co}$. Then, observation of
LiH at the mJy level would require the USO to be young, and might mean
a significantly small value of $S_{\rm lih}$ as long as CO is not
detected. How transient is the LiH gas phase abundance likely to be?
The dynamical evolution time of a USO is about a free-fall time, which
is about 5.75 $\times 10^8$ s for our standard case of $M_{\rm
J}$. Even if the USOs are more long-lived than the dynamical time, the
gas and dust would circulate at the sound speed on this time-scale,
and this circulation should ensure a fresh supply of gas-phase LiH and
CO as the dust from the cloud core is exposed to a less protective
environment.  Thus if we were to observe about 100 USOs, we would
expect at least one of them to emit a strong LiH line and a
supplementary weak CO line.  This is performed simply by observing the
brightest USOs.

Since LiH is transient, it is useful to calculate the emission flux with
a smaller abundance than that of the typical model with a Jupiter
mass. We try to estimate this by postulating abundances of $10^{-12}$
and $10^{-14}$.  The expected line fluxes are 4.81 mJy for $10^{-12}$
and 0.35 mJy for $10^{-14}$.  At the abundance of $10^{-12}$, the
emission line is still optically thick, and the flux hardly decreases
relative to that of the typical case. In the case of $10^{-14}$, the
emission line is optically thin, and then the flux is reduced. The
time-scale for the decrement from $10^{-10}$ to $10^{-14}$ is only
several times the depletion timescale. Thus, the detected USOs are
expected to be younger than the dynamical time-scale, as long as the
spallation of cosmic-ray particles on nuclei at the surface layesr of
USOs is not efficient. Furthermore, if such very low abundance molecule
is observable, $^6$LiH may be detectable near LiH by ALMA as long as the
USO is very young. This is because abundance of $^6$Li can be about
$10^{-13}$ even if it is primordial.  When a USO is old, abundance of
$^6$LiH decreses much below $10^{-14}$. We may find $^6$LiH line near
LiH, when we observe more USOs than a hundred.

We also comment on the dynamical coupling between gas and dust.  We
consider the typical case of a cloud mass of $M_J$.  The
time-scale for dust interaction with the gas is $1/(\pi a^2 n_{\rm
dust} v_{\rm s})$, and is equal to 3.3 $\times 10^5$ s if a standard
MNR distribution of dust is assumed. The dynamical timescale, which
can describe the evolution timescale, is about 5.75 $\times 10^8$ s,
the free-fall timescale.  Thus the dust frequently interacts with gas
during the dynamical evolution of a galactic USO.  Hence, the coupling
between dust and cold gas can be well established on the dynamical
time-scale.  Since the dust-gas coupling is good, this also means that
the temperature of dust and gas is nearly the same. That is, the
effective specific heat ratio can be smaller than 4/3. Hence, the
expected cold cloud should be transient on a dynamical time-scale
since it is unstable to gravitational contraction. This is another
constraint on a model for USOs.  The temperature of the clouds may be
eventually confirmed by detection of CO lines if future high
resolution of spectroscopy is possible (Sciama 2000). 

The preceding discussion does not contradict the universality of the
galactic USOs, although the observed USOs should be transient chemically
and dynamically.  However we need to find a production process for
transient USOs.  An interesting recent discussion proposes an origin for
free-floating compact substellar objects of about Jupiter mass (Boss
2001). According to the paper by Boss, searches for very low mass
objects in young star clusters have uncovered evidence for free-floating
objects with inferred masses possibly as low as 5-15 Jupiter masses,
similar to the masses of several extrasolar planets. He shows that the
process that forms single and multiple protostars, namely, collapse and
fragmentation of molecular clouds, might be able to produce
self-gravitating objects with initial masses less than 1 $M_J$, provided
that magnetic field tension effects are important and can be represented
approximately by diluting the gravitational field. If such fragments can
be ejected from an unstable quadruple protostar system, prior to gaining
significantly more mass, protostellar collapse might then be able to
explain the formation of free-floating objects with masses below 13
$M_J$.  We note that ejected material can be in the form of extremely
dense and cold clumps, and speculate that this mechanism may be
responsible to the origin of a galactic population of USOs.  A more
precise examination of this possibility is needed.

In any case, if LiH and CO were to be detected from galactic USOs, it
would be necessary to consider a production source for the USOs. A
promising candidate mechanism is the gravito-magnetic scattering of
dense clumps from star-forming regions (Boss 2001).  In the galactic
halo, stars can form, for example, in HI supershells and in infalling,
compressed HI high velocity clouds.  Hence free-floating USOs may
originate in halo star-forming regions, although any contribution to
the overall population of such objects is highly uncertain because of
our sparse knowledge of halo star-forming regions.

\section{Conclusions}

It has been suggested that some halo and disk dark matter could reside
in the form of cold dark clouds and so be undetected. In particular,
Lawrence (2001) has constrained the physical properties of such halo
baryonic dark matter candidate objects by appealing to submm-mm
observations. According to our observational predictions for Lawrence's
objects, LiH emission lines can in principle be detected by ALMA. This
will provide a strong constraint on models for the galactic
unidentified SCUBA sources, which we refer to as USOs in this $Letter$. 
According to Gerhard \& Silk (1996), galactic USOs can be dynamically
transient. In the current work, furthermore, since the chemical
depletion of CO and presumably LiH is significant, then they are also
chemically transient. Hence, we may need to observe around 100 USOs for
clear evidence of the galactic USO candidates that we have
postulated. If their reality is established, the contribution of USOs
as well as MACHOs to the galactic baryon budget will need to be
re-examined.

%
%

\section*{ACKNOWLEDGEMENT}
We express our gratitude for the referee, A. Lawrence, whose comments
helped clarify the discussions and presentation of the manuscript.  H.K. 
is grateful to Profs. S.Mineshige, S.Inagaki and R.Hirata for their
encouragement.

%
%


\begin{table*}
\caption{Expected Emission Lines} 
\begin{tabular}{crrr}
\hline 
\hline
--- & $0.1M_{\rm J}$ &  $1.0M_{\rm J}$ &  $10.0M_{\rm J}$ \\ 
\hline
Radius (AU) & 1.0  & 10.0  & 100  \\
Density (cm$^{-3}$) & $4.0\times 10^{12}$ & $4.0\times 10^{10}$  &
 $4.0\times 10^{8}$  \\
Population mass ($M_\odot$ pc$^{-3}$) & 0.3 & 0.01 & 0.001 \\
Distance (pc) & 45  & 300  & 1000  \\
\hline
HD (mJy) & 5.3 10$^{-3}$ & 1.2 10$^{-3}$ &  1.21 10$^{-3}$ \\
LiH (mJy) & 2.1  & 4.8 &  43.3 \\
CO (mJy) & 3.6 10$^{-2}$ & 8.2 10$^{-2}$ &  73.0 10$^{-2}$ \\
dust cont. at 0.11 mm (mJy) & 3.9 10$^{-6}$ &8.8 10$^{-6}$  &7.9 10$^{-5}$\\
dust cont. at 0.67 mm (mJy) & 0.075 &0.17  &  1.52 \\
dust cont. at 2.60 mm (mJy) & 2.2 10$^{-2}$ &4.9 10$^{-2}$ & 0.44 \\
\hline
\end{tabular}
\end{table*}


\begin{figure}
\begin{center}
\includegraphics[height=10cm,clip,keepaspectratio]{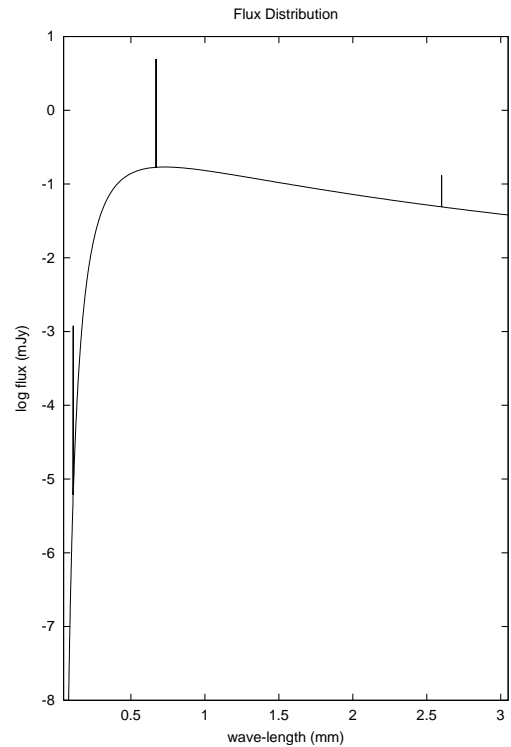}
\caption{
Schematic spectral of a typical galactic USO of $M_{\rm J}$. The
continuum is optically thick dust black body. Each of three lines is
HD (0.11 mm), LiH (0.67 mm), and CO (2.60 mm), respectively. The
vertical axis is ticked in log-scale of mJy, while the horizontal axis
in normal wavelength of mm.
}
\end{center}
\end{figure}

\end{document}